# Desensitized Kalman Filtering with Analytical Gain

Taishan Lou

*School of Electric and Information Engineering, Zhengzhou University of Light Industry, Zhengzhou, 45002, China,*

*tayzan@sina.com*

**Abstract:** The possible methodologies to handle the uncertain parameter are reviewed. The core idea of the desensitized Kalman filter is introduced. A new cost function consisting of a posterior covariance trace and trace of a weighted norm of the state error sensitivities matrix is minimizing to obtain a well-known analytical gain matrix, which is different from the gain of the desensitized Kalman filter. The pre-estimated uncertain parameter covariance is set as a referential sensitivity-weighting matrix in the new framework, and the rationality and validity of the covariance are tested. Then, these results are extended to the linear continuous system.

## I. Introduction

The Kalman filter can obtain optimal estimation based on a fundamental assumption that the dynamic models can be accurately modeled without any colored noise or uncertain parameters. However, in practice, these models always include many additional parameters, whose uncertainties always result in the poor state estimate.

Recent literature in optimal estimation theory plays more interesting in mitigating these negative effects of the parameter uncertainty [1-6]. There are four possible methodologies to handle the parameter uncertainty problem. The first one is to completely neglect uncertain parameter. Ignoring low impact or well-calibrated parameter is reasonable, but ignoring high impact parameter uncertainty can bring large bias errors in the estimated state. The second one is to expand the state vector to include the parameters that may be uncertain as additional states. This method will lead to considerable computational power and processing time required, especially for the large dimension systems. The third one is to "consider" the parameters, which is known as the Schmidt-Kalman filter [1] or consider Kalman filter[2]. This method is that the state and covariance estimate are updated by using the pre-estimated parameter covariance, without estimating these parameters directly. This approach decreases the cost of the computational power and processing time required comparison to the second. One drawback is that it requires knowledge of covariance of the parameter uncertainties. The fourth one is to decrease the sensitive to deviations, which includes the robust filters and the recent presented desensitized Kalman filter by minimizing a cost function augmented by a penalty function [6,7]. These robust filters always need the norm-bounded parameter uncertainty. The desensitized Kalman filter proposed by Karlgaard and Shen is another type of the robust Kalman filter with knowledge of the sensitivity-weighting matrix [6].

Desensitized Kalman filter (DKF) was first developed by Karlgaard and Shen as means to account for the model parameter uncertainties by using desensitized optimal control technique in reference [6]. They penalized the cost function consisting of the posterior covariance trace by a weighted norm of the state error sensitivities, which means that this cost function was augmented with the penalty function consisting of a weighted norm of the state error sensitivities. Desensitized state estimates were obtained by minimizing the above cost function. Then, they extended the concept of the DKF to desensitized unscented Kalman filtering [8], desensitized divided difference filtering [9], in which the cost function was augmented the same penalty function. The DKF is non-minimum variance, but exhibits reduced sensitivity to deviations in the assumed dynamic model parameters. The DKF was applied into an induction motor state estimation problem with parameter uncertainties [10], and the effectiveness of DKF was demonstrated. However, the DKF has two disadvantages over the conventional Kalman filter. The first is to known the sensitivity-weighting matrix. The second is to that obtain the gain matrix only by solving a linear equation without an analytical solution. When the dimension of the state vector or parameter vector is large, the cost of the computational power and processing time required increase sharply.

This note recombines the state error sensitivities vector of each parameter to a total sensitivity matrix of all parameters in the condition of the linear discrete model. A new cost function is the sum of the posterior covariance trace and the trace of a weighted norm of the state error sensitivities matrix, not the sum of each penalty function in literature . Minimizing this new cost function gives an analytical solution of the gain matrix, which has the same structure with the conventional Kalman filter. Based on this new framework, we extend the results to the linear continuous model.

## II. Desensitized Linear Kalman Filter with Analytical Gain

2.1 Desensitized Discrete Linear Kalman Filter with Analytical Gain

Without loss of generality, we discard the deterministic input and some matrices in the linear discrete model in reference [7] to simplify the algorithm. Consider the process and measurement models given by

$$\mathbf{x}_k = \mathbf{\Phi}_{k-1}(\mathbf{p})\mathbf{x}_{k-1} + \mathbf{w}_{k-1} \tag{1}$$

$$\mathbf{z}_k = \mathbf{H}_k(\mathbf{p})\mathbf{x}_k + \mathbf{v}_k \tag{2}$$

where $\mathbf{x}_k$ is the $n \times 1$ state vector, and $\mathbf{z}_k$ is the $m \times 1$ measurement vector. $\mathbf{\Phi}_{k+1|k}$ is the state transition matrices, $\mathbf{H}_k$ is the measurement matrix. $\mathbf{p}$ is referred to as the $\ell \times 1$ uncertain parameter vector. $\mathbf{w}_k$ and $\mathbf{v}_k$ are independent zero-mean Gaussian noise processes, and their covariance are respectively $\mathbf{Q}_k$ and $\mathbf{R}_k$. They satisfy

$$E[\mathbf{w}_i \mathbf{w}_j^T] = \mathbf{Q}_k \delta_{ij}, \ E[\mathbf{v}_i \mathbf{v}_j^T] = \mathbf{R}_k \delta_{ij}, \ E[\mathbf{w}_i \mathbf{v}_j^T] = \mathbf{0} \tag{3}$$

where $\delta_{ij}$ is the Kronecker delta function, and $\mathbf{Q}_k > 0, \mathbf{R}_k > 0$.

In this work, the uncertain model parameter is given an estimate, $\mathbf{p} = \hat{\mathbf{p}}$, with the a priori knowledge. In the Kalman filter, the estimated state propagation equation

$$\hat{\mathbf{x}}_k^- = \overline{\mathbf{\Phi}}_{k-1} \hat{\mathbf{x}}_{k-1}^+ \tag{4}$$

and the measurement updated equation

$$\hat{\mathbf{z}}_k^- = \overline{\mathbf{H}}_k \hat{\mathbf{x}}_k^- \tag{5}$$

where the superscripts $^-$ denote a priori and $^+$ denote a posteriori, and the overbar indicates the corresponding estimate function of the parameter, such as $\overline{\mathbf{\Phi}} = \mathbf{\Phi}(\hat{\mathbf{p}})$. Then, the Kalman filter provides an optimal blending of the $\hat{\mathbf{x}}_k^-$ and $\mathbf{z}_k$ to obtain the a posteriori state estimate via

$$\hat{\mathbf{x}}_k^+ = \hat{\mathbf{x}}_k^- + \mathbf{K}_k (\mathbf{z}_k - \hat{\mathbf{z}}_k^-) \tag{6}$$

We define the a priori estimation error as $\mathbf{e}_k^- = \hat{\mathbf{x}}_k^- - \mathbf{x}_k$ and the a posteriori estimation error as $\mathbf{e}_k^+ = \hat{\mathbf{x}}_k^+ - \mathbf{x}_k$, and assume these estimation errors to be zero-mean. Then, the associated error covariance matrices are

$$\mathbf{P}_k^- = E[\mathbf{e}_k^- \mathbf{e}_k^{-T}] = \overline{\mathbf{\Phi}}_{k-1} \mathbf{P}_{k-1}^+ \overline{\mathbf{\Phi}}_{k-1}^T + \mathbf{Q}_{k-1} \tag{7}$$

$$\mathbf{P}_k^+ = E[\mathbf{e}_k^+ \mathbf{e}_k^{+T}] = (\mathbf{I} - \mathbf{K}_k \overline{\mathbf{H}}_k) \mathbf{P}_k^- (\mathbf{I} - \mathbf{K}_k \overline{\mathbf{H}}_k)^T + \mathbf{K}_k \mathbf{R}_k \mathbf{K}_k^T \tag{8}$$

which is valid for any $\mathbf{K}_k$. The Kalman optimal gain $\mathbf{K}_k$ is chosen by minimizing the cost function $J = Tr(\mathbf{P}_k^+)$, and it is given by

$$\mathbf{K}_k = \mathbf{P}_k^- \bar{\mathbf{H}}_k^T \boldsymbol{\Xi}_k^{-1} \tag{9}$$

where $\boldsymbol{\Xi}_k = \bar{\mathbf{H}}_k \mathbf{P}_k^- \bar{\mathbf{H}}_k^T + \mathbf{R}_k$, and "*Tr*" denotes the trace of the matrix.

Under the fundamental assumptions of the Kalman filter (zero-mean white-noise sequence, unbiased a priori estimation errors, no model and parameter uncertainty, known process and measurement models, etc.), the state estimate and state estimation error covariance updates are optimal. In the presence of model parameter uncertainties, the dynamic model cannot match the true model. This means that the fundamental assumptions of the Kalman filter cannot be satisfied, and the state estimates may be biased and even divergence. Karlgaard and Shen [7] proposed a desensitized optimal filtering to mitigate the negative effects of the uncertain parameters based on the cost function of the state error sensitivities. They defined the state error sensitivities and propagation equations of each parameter component $p_i$ of $\mathbf{p}$ as

$$\boldsymbol{\sigma}_{i,k}^- = \frac{\partial \mathbf{e}_k^-}{\partial p_i} = \frac{\partial \hat{\mathbf{x}}_k^-}{\partial p_i} = \bar{\boldsymbol{\Phi}}_{k-1} \boldsymbol{\sigma}_{i,k-1}^+ + \frac{\partial \bar{\boldsymbol{\Phi}}_{k-1}}{\partial p_i} \hat{\mathbf{x}}_{k-1}^+ \tag{10}$$

$$\boldsymbol{\sigma}_{i,k}^+ = \frac{\partial \mathbf{e}_k^+}{\partial p_i} = \frac{\partial \hat{\mathbf{x}}_k^+}{\partial p_i} = \boldsymbol{\sigma}_{i,k}^- - \mathbf{K}_k \boldsymbol{\gamma}_k \tag{11}$$

where $\boldsymbol{\gamma}_k = \bar{\mathbf{H}}_k \boldsymbol{\sigma}_{i,k}^- + \frac{\partial \bar{\mathbf{H}}_k}{\partial p_i} \hat{\mathbf{x}}_k^-$. Note that the sensitivity of the true state is $\partial x_k / \partial p_i = 0$ in above formulations, and it is assumed that $\partial \mathbf{K}_k / \partial p_i = 0$ in Eq. (11).

A cost function consisting of the posterior covariance and a weighted norm of the posterior sensitivity is proposed as

$$J = Tr(\mathbf{P}_k^+) + \sum_{i=1}^{\ell} \boldsymbol{\sigma}_{i,k}^{+T} \mathbf{W}_i \boldsymbol{\sigma}_{i,k}^+ \tag{12}$$

where $\mathbf{W}_i$ is a $n \times n$ symmetric positive semi-definite weighting matrix for the *i*th sensitivity. Then, taking the derivative with respect to $\mathbf{K}_k$, using the trace derivative properties found in Appendix A, and setting $\partial J / \partial \mathbf{K}_k = 0$, yields

$$\mathbf{K}_k \mathbf{\Xi}_k + \sum_{i=1}^{\ell} \mathbf{W}_i \mathbf{K}_k \boldsymbol{\gamma}_{i,k} \boldsymbol{\gamma}_{i,k}^T = \mathbf{P}_k^- \bar{\mathbf{H}}_k^T + \sum_{i=1}^{\ell} \mathbf{W}_i \boldsymbol{\sigma}_{i,k}^- \boldsymbol{\gamma}_{i,k}^T \tag{13}$$

Note that the gain $\mathbf{K}_k$ must be solved with the linear equation in Eq. (13) differing the analytical gain matrix in the conventional Kalman filter. This implies that the cost of the computational power and the processing time required will increase rapidly, especially when the dimension of the state is large.

In this work, we redefine the cost function, which also consists of the posterior covariance and another weighted norm of the posterior sensitivity, and obtain an analytical solution of the gain matrix. We redefine the state error sensitivities and propagation equations of the parameter vector $\mathbf{p}$ as

$$\mathbf{S}_k^- = \frac{\partial \mathbf{e}_k^-}{\partial \mathbf{p}} = \frac{\partial \hat{\mathbf{x}}_k^-}{\partial \mathbf{p}} = \bar{\mathbf{\Phi}}_{k-1} \mathbf{S}_{k-1}^+ + \bar{\mathbf{\Psi}}_{k-1} \tag{14}$$

$$\mathbf{S}_k^+ = \frac{\partial \mathbf{e}_k^+}{\partial \mathbf{p}} = \frac{\partial \hat{\mathbf{x}}_k^+}{\partial \mathbf{p}} = \mathbf{S}_k^- - \mathbf{K}_k \boldsymbol{\gamma}_k \tag{15}$$

where

$$\bar{\mathbf{\Psi}}_{k-1} = \frac{\partial \bar{\mathbf{\Phi}}_{k-1}}{\partial \mathbf{p}} \hat{\mathbf{x}}_{k-1}^+ . \tag{16}$$

$$\boldsymbol{\gamma}_k = \bar{\mathbf{H}}_k \mathbf{S}_k^- + \bar{\mathbf{H}}_k^p . \tag{17}$$

where $\bar{\mathbf{H}}_k^p = \frac{\partial \bar{\mathbf{H}}_k}{\partial \mathbf{p}} \hat{\mathbf{x}}_k^-$.

We redefine a new cost function based on the trace of the weighted norm of the posterior sensitivity matrix given by

$$J_a = Tr(\mathbf{P}_k^+) + Tr(\mathbf{S}_k^+ \mathbf{W}_a \mathbf{S}_k^{+T}) \tag{18}$$

where $\mathbf{W}_a$ is a $\ell \times \ell$ symmetric positive semi-definite weighting matrix for the uncertain parameters. Substituting Eqs. (8) and (15) into Eq. (18) and taking the derivative with respect to the gain $\mathbf{K}_k$, and using the trace derivative properties in Appendix A, yields

$$\frac{\partial J_a}{\partial \mathbf{K}_k} = \frac{\partial}{\partial \mathbf{K}_k} Tr(\mathbf{P}_k^+) + \frac{\partial}{\partial \mathbf{K}_k} Tr(\mathbf{S}_k^+ \mathbf{W}_a \mathbf{S}_k^{+T})$$
$$= 2\mathbf{K}_k \mathbf{\Xi}_k - 2\mathbf{P}_k^- \bar{\mathbf{H}}_k^T - 2\mathbf{S}_k^- \mathbf{W}_a \boldsymbol{\gamma}_k^T + 2\mathbf{K}_k \boldsymbol{\gamma}_k \mathbf{W}_a \boldsymbol{\gamma}_k^T \qquad (19)$$

Setting $\partial J_a / \partial \mathbf{K}_k = 0$ and simplifying the formulation gives the analytical gain matrix as follows

$$\mathbf{K}_k = (\mathbf{P}_k^- \bar{\mathbf{H}}_k^T + \mathbf{S}_k^- \mathbf{W}_a \boldsymbol{\gamma}_k^T)(\mathbf{\Xi}_k + \boldsymbol{\gamma}_k \mathbf{W}_a \boldsymbol{\gamma}_k^T)^{-1} \qquad (20)$$

Note that the formulation of the gain $\mathbf{K}_k$ in Eq. (20) is the same as the conventional Kalman filter in form and it is an analytical solution, too. Corresponding to the method in reference [6,7], the proposed cost function and gain formulation are completeness and clear at a glance, and the cost of the computational power and the processing time required decrease greatly. Even more important, it provides a new algorithm framework for the desensitized optimal filtering including the discrete linear filter and the continuous linear filter.

2.2 Desensitized Continuous Linear Kalman Filter with Analytical Gain

In this study, the corresponding result for the linear continuous model with the new cost function and the analytical gain in the new framework is summarized. The KSDKF for the continuous case is in Appendix B.

Consider the continuous linear system

$$\dot{\mathbf{x}}(t) = \mathbf{\Phi}(\mathbf{p},t)\mathbf{x}(t) + \mathbf{w}(t) \qquad (21)$$

$$\mathbf{z}(t) = \mathbf{H}(\mathbf{p},t)\mathbf{x}(t) + \mathbf{v}(t) \qquad (22)$$

where $\mathbf{x}(t)$ is the $n \times 1$ state vector, and $\mathbf{z}(t)$ is the $m \times 1$ measurement vector. $\mathbf{\Phi}(\mathbf{p},t)$ and $\mathbf{H}(\mathbf{p},t)$ are the state transition matrix and the measurement matrix. $\mathbf{w}(t)$ and $\mathbf{v}(t)$ satisfy $\mathbf{w} \sim N(0, \mathbf{Q}(t)\delta(t-\tau))$ and $\mathbf{v} \sim N(0, \mathbf{R}(t)\delta(t-\tau))$.

The state estimate error is define as $\mathbf{e}(t) = \hat{\mathbf{x}}(t) - \mathbf{x}(t)$. Then, the sensitivity of the error to the parameter vector $\mathbf{p}$ is

$$\mathbf{S} = \frac{\partial \mathbf{e}(t)}{\partial \mathbf{p}} = \frac{\partial \hat{\mathbf{x}}(t)}{\partial \mathbf{p}} \quad (23)$$

and the corresponding sensitivities obey the propagation equation

$$\dot{\mathbf{S}} = \bar{\mathbf{\Phi}} \mathbf{S} + \frac{\partial \bar{\mathbf{\Phi}}}{\partial \mathbf{p}} \hat{\mathbf{x}} - \mathbf{K} \boldsymbol{\gamma} \quad (24)$$

where $\bar{\mathbf{\Phi}} = \mathbf{\Phi}(\hat{\mathbf{p}}, t)$, $\bar{\mathbf{H}} = \mathbf{H}(\hat{\mathbf{p}}, t)$ and $\boldsymbol{\gamma} = \bar{\mathbf{H}} \mathbf{S} + \frac{\partial \bar{\mathbf{H}}}{\partial \mathbf{p}} \hat{\mathbf{x}}$.

The new cost function, which is reformulated to minimize the rate of change of the state error covariance, augmented by the new penalty function is

$$J_a = Tr(\dot{\mathbf{P}}) + 2Tr(\mathbf{S} \mathbf{W}_a \dot{\mathbf{S}}^T) \quad (25)$$

In the continuous filter case, the optimal gain is obtained as the discrete filter, which is found by taking the derivative with respect to the gain $\mathbf{K}$, and the result is

$$\mathbf{K} = (\mathbf{P} \bar{\mathbf{H}}^T + \boldsymbol{\gamma} \mathbf{W}_a \mathbf{S}^T) \mathbf{R}^{-1} \quad (26)$$

2.3 Remarks

**Remark 1**: The penalty functions in the two algorithms for the discrete case are different, because the KSDKF considers the sensitivity to each parameter respectively and the ADKF does this as a whole. From their definitions, they satisfy

$$\mathbf{S}_k^- = (\boldsymbol{\sigma}_{1,k}^-, \boldsymbol{\sigma}_{2,k}^-, \cdots, \boldsymbol{\sigma}_{\ell,k}^-)_{n \times \ell} \quad (27)$$

$$\mathbf{S}_k^+ = (\boldsymbol{\sigma}_{1,k}^+, \boldsymbol{\sigma}_{2,k}^+, \cdots, \boldsymbol{\sigma}_{\ell,k}^+)_{n \times \ell} \quad (28)$$

The aforementioned two different definitions generate the two different penalty functions, $\sum_{i=1}^{\ell} \boldsymbol{\sigma}_{i,k}^{+T} \mathbf{W}_i \boldsymbol{\sigma}_{i,k}^+$ and $Tr(\mathbf{S}_k^+ \mathbf{W}_a \mathbf{S}_k^{+T})$. The optimal gain $K_k$ of the KSDKF in Eq. (13) is obtained by solve a linear matrix equation, and the one of the ADKF in Eq. (20) is founded as the conventional

Kalman filter in a well-known form and it also is a closed-form solution. The optimal gain of the ADKF greatly decrease the computational power cost and the processing time required.

**Remark 2**: The relations of the two sensitivity definitions are the same as the discrete case. The definition of the penalty function of the KSDKF in Appendix B, which is different from the discrete case, is the product of the state error sensitivity, its rate of change and the corresponding sensitivity matrix. Because the state error sensitivity has not the gain matrix, so the analytical gain is obtained by taking the derivative with respect to the gain. So is in the ADKF for the continuous case.

**Remark 3**: The sensitivity-weighting matrix $\mathbf{W}_i$ in Eq. (12) is a $n \times n$ weighting matrix for the $i^{\text{th}}$ sensitivity respect to the state. But, how to determine each sensitivity-weighting matrix in Eq. (12) is not proposed in reference [6,7], and is an open problem to be resolved. The sensitivity-weighting matrix $\mathbf{W}_a$ in Eq. (18) is a $\ell \times \ell$ weighting matrix respect to the uncertain parameters. Considering the roles of the pre-estimated uncertain parameter covariance as in the consider Kalman filter [2,11]. This covariance may be chosen as a referential sensitivity-weighting matrix in ADKF. It is reasonable that assigning each covariance of the uncertain parameter to the corresponding sensitivity-weighting respect to all the states as in the consider Kalman filter. Then, we demonstrate this in the following numerical simulations.

**Remark 4**: The gain formulation of ADEKF makes the DEKF recover the well-known form of the Kalman filter, in which the filtering algorithm mainly includes five basic equations. Even more important, it provides a new algorithm framework for the DEKF including the discrete nonlinear model, the continuous nonlinear model and the mixed continuous-discrete nonlinear model.

## III. Numerical Results

To compare the performance of the proposed ADKF and the KSDKF, the linear discrete-time dynamic stochastic system in literature [6] is considered. The corresponding dynamic and measurement equations are

$$\mathbf{x}_{k+1} = \begin{bmatrix} 1 & 0.1+\alpha \\ \beta-0.5 & 0.9 \end{bmatrix} \mathbf{x}_k + \mathbf{w}_k \tag{29}$$

$$\mathbf{z}_k = \mathbf{x}_k + \mathbf{v}_k \tag{30}$$

where $\alpha$ and $\beta$ are two scalar uncertain model parameters, which are assumed to be constants that $\alpha \sim U(-0.1, 0.1)$ and $\beta \sim U(-0.5, 0.5)$, independently. As in literature [6], we assumed that the initial state $\mathbf{x}_0 = [10, -10]^T$ and $\mathbf{P}_0 = 0.1\mathbf{I}_{2\times 2}$. The noises are $\mathbf{w}_k \sim N(\mathbf{0}_{2\times 1}, 0.1\mathbf{I}_{2\times 2})$ and $\mathbf{v}_k \sim N(\mathbf{0}_{2\times 1}, \mathbf{I}_{2\times 2})$.

In this work, we only consider two sets of values of the sensitivity-weighting matrix in an effort to compare these two methods. The sensitivity-weight matrix of the ADKF is set as $\mathbf{W}_a = diag[0.003, 0.075]$, which is ninety percent of the true covariance of the two uncertain parameters; the sensitivity-weight matrices of the KSDKF have two values, $\mathbf{W}_1 = \mathbf{W}_2 = diag[0.003, 0.075]$ and $\mathbf{W}_1 = \mathbf{W}_2 = 0.1\mathbf{I}_{2\times 2}$, which is in literature [6]. Two 5000 case Monte-Carlo simulations with 50 samples for the two filters runs are performed. In additional, the nominal values for the uncertain parameters are $\hat{\alpha} = 0, \hat{\beta} = 0$. The root mean square (RMS) errors of the state estimate and the penalty and total cost functions are calculated to compare the performance of the two methods at each epoch.

Figures 1 and 2 show the results when these sensitivity-weighting matrices are made equal as in the first set. It can be seen that the RMS errors of the first state $x_1$ of the ADKF are better than the KSDKF in Fig.1 (a), and the RMS errors of the second state $x_2$ are about the same for the two methods in Fig.1 (b). Figure 2 shows that the cost/penalty functions of the ADKF are all slightly smaller than these of the KSDKF are. In a word, the performance of the ADKF is better than the KSDKF when they have the same sensitivity-weighting matrices.

When these sensitivity-weighting matrices are set as in the second set, Figures 3 and 4 show the results. Fig.3 (a) shows that the RMS errors of the first state $x_1$ of the ADKF are better than the KSDKF, and in Fig.3 (b) the RMS errors of the second state $x_2$ have almost identical performance for the two methods. The cost and penalty functions of the ADKF are all smaller than the KSDKF in Fig.4. This is maybe because the first parameter sensitivity-matrix of the KSDKF is not suitable, which is one order greater in magnitude than that of the ADKF. Therefore, it is reasonable to choose the a priori covariance of the uncertain parameters as the sensitivity-weighting matrix in Eq. (18).

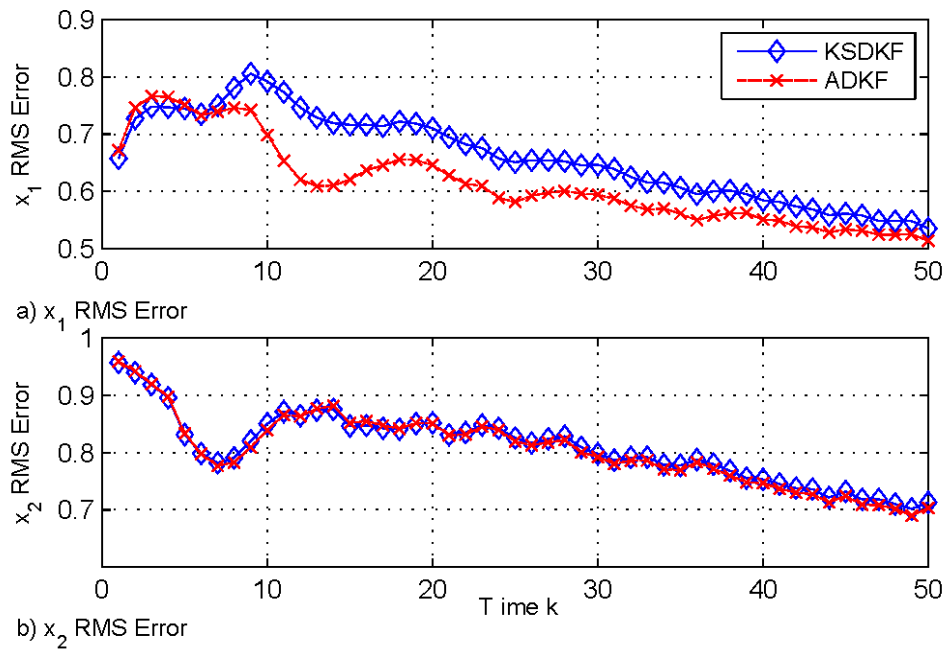

**Fig. 1 State RMS errors for the second set values.**

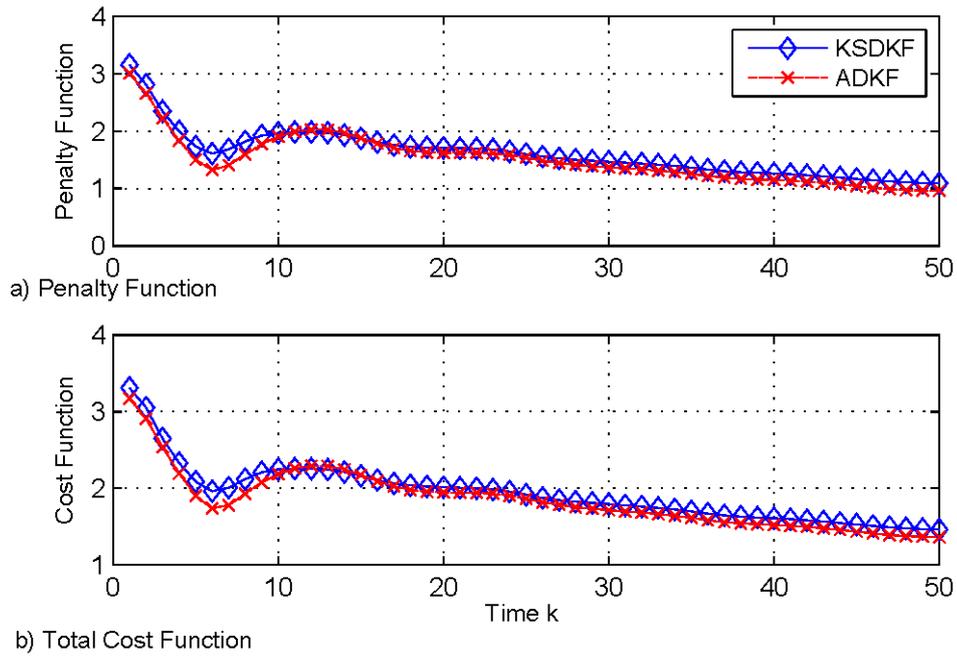

a) Penalty Function

b) Total Cost Function

**Fig. 2 Cost function and penalty function for the second set values.**

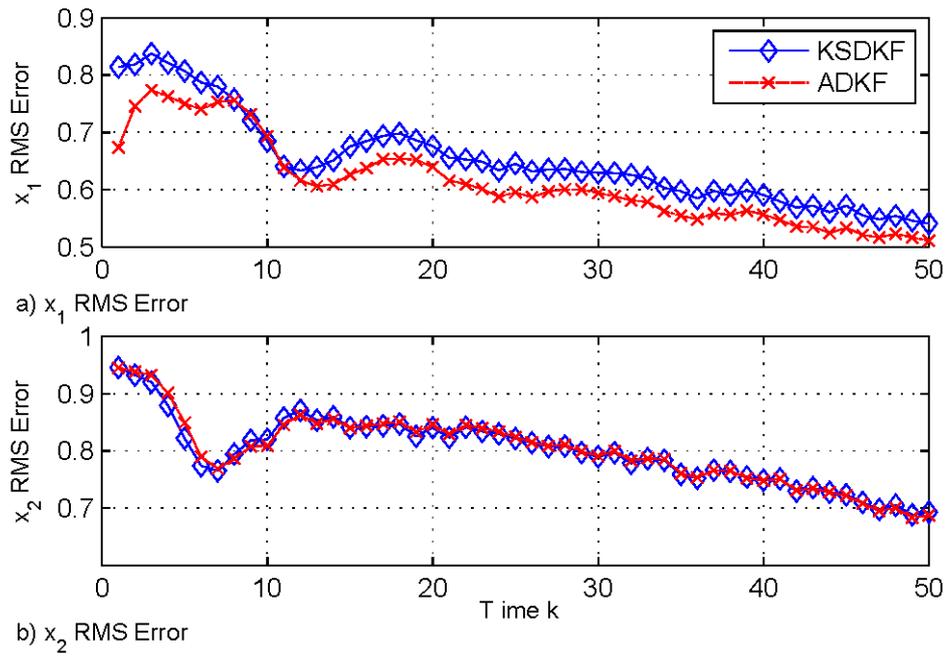

a) $x_1$ RMS Error

b) $x_2$ RMS Error

**Fig. 3 State RMS errors for the third set values.**

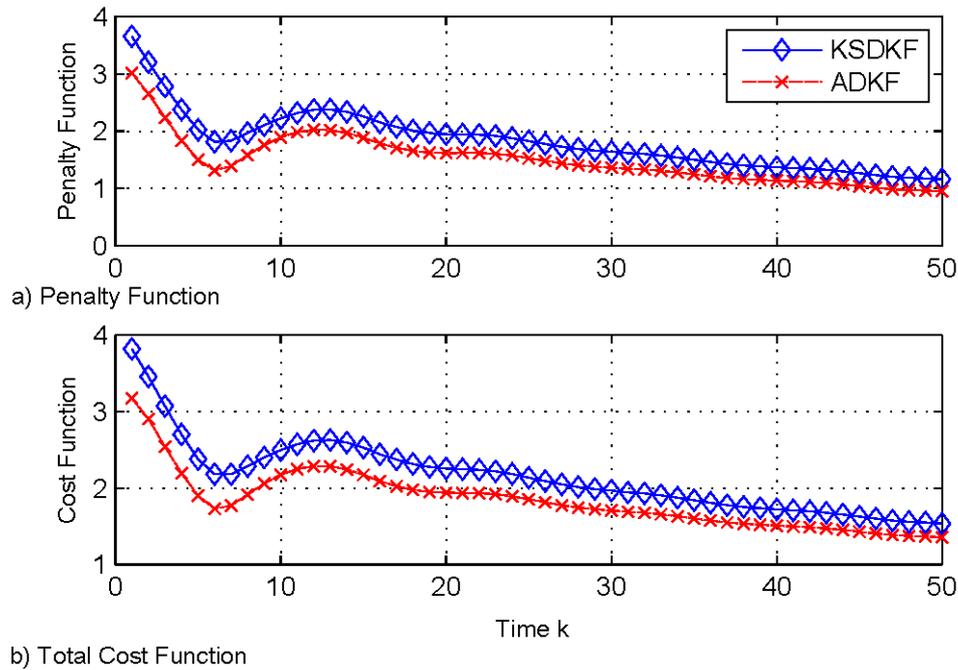

a) Penalty Function

b) Total Cost Function

**Fig. 4 Cost function and penalty function for the third set values.**

## Appendix A- Matrix Trace Calculus

To get the optimal gain from the cost function in Kalman filter derivations, taking the partial derivative of the trace of matrix is often used. The corresponding results about the derivatives are

$$\frac{\partial}{\partial \mathbf{K}} Tr(\mathbf{KP}) = \mathbf{P}^T \tag{31}$$

$$\frac{\partial}{\partial \mathbf{K}} Tr(\mathbf{PK}^T) = \mathbf{P} \tag{32}$$

$$\frac{\partial}{\partial \mathbf{K}} Tr(\mathbf{KPK}^T) = \mathbf{KP}^T + \mathbf{KP} \tag{33}$$

Where $\mathbf{K}$ and $\mathbf{P}$ are two arbitrary matrices satisfying matrix multiplication rules.

## Appendix B- Desensitized Continuous Linear Kalman Filter

In this appendix, the desensitized continuous linear Kalman filter, which was proposed by Karlgaard and Shen [6], is reviewed. Consider the continuous linear system in Eqs. (21) and (22), the linear Kalman filter for the continuous case is

$$\dot{\hat{\mathbf{x}}}(t) = \bar{\mathbf{\Phi}}\hat{\mathbf{x}}(t) + \mathbf{K}\{\mathbf{z}(t) - \bar{\mathbf{H}}\hat{\mathbf{x}}(t)\} \tag{34}$$

$$\dot{\mathbf{P}} = (\bar{\mathbf{\Phi}} - \mathbf{K}\bar{\mathbf{H}})\mathbf{P} + \mathbf{P}(\bar{\mathbf{\Phi}} - \mathbf{K}\bar{\mathbf{H}})^T + \mathbf{Q} + \mathbf{KWK}^T \tag{35}$$

Then, the sensitivity of the error to the parameter $p_i$ is

$$\hat{\mathbf{\sigma}}_i = \frac{\partial \mathbf{e}}{\partial p_i} = \frac{\partial \hat{\mathbf{x}}(t)}{\partial p_i} \tag{36}$$

and the corresponding sensitivity propagation equation is

$$\dot{\hat{\mathbf{\sigma}}}_i = \bar{\mathbf{\Phi}}\hat{\mathbf{\sigma}}_i + \frac{\partial \bar{\mathbf{\Phi}}}{\partial p_i}\hat{\mathbf{x}}(t) - \mathbf{K}\mathbf{\gamma}_i \tag{37}$$

where $\mathbf{\gamma}_i = \bar{\mathbf{H}}\hat{\mathbf{\sigma}}_i + \frac{\partial \bar{\mathbf{H}}}{\partial p_i}\hat{\mathbf{x}}(t)$.

The cost function of the conventional Kalman filter, $J = Tr(\dot{\mathbf{P}})$, is augmented by a penalty function, which is the product of the state error sensitivity, its rate of change and the corresponding sensitivity matrix. The augmented cost function is

$$J = Tr(\dot{\mathbf{P}}) + 2\sum_{i=1}^{\ell}\hat{\boldsymbol{\sigma}}_i^T \mathbf{W}_i \dot{\hat{\boldsymbol{\sigma}}}_i \tag{38}$$

The optimal gain is obtained by taking the derivative with respect to the gain $\mathbf{K}$, and the corresponding result is

$$\mathbf{K} = (\mathbf{P}\bar{\mathbf{H}}^T + \sum_{i=1}^{\ell} \mathbf{W}_i \hat{\boldsymbol{\sigma}}_i \boldsymbol{\gamma}_i^T)\mathbf{R}^{-1} \tag{39}$$